\def\aj{{AJ}}

\def\apj{{ApJ}}

\def\mnras{{MNRAS}}

\documentclass[cup5b]{caps}

\newcommand{\mcmc}        {M_{\rm CMC}}
\newcommand{\ecmc}        {\epsilon_{\rm CMC}}
\newcommand{\tg}          {t_{\rm growth}}
\newcommand{\mdisk}        {M_{\rm disk}}
\newcommand{\dt}          {\Delta t}

\newcommand{\etal}{\textrm{et~al. }}
\newcommand{\ie}{\textrm{i.e. }}
\newcommand{\eg}{\textrm{e.g. }}

\catcode`\@=11 
\def\la{\mathrel{\mathpalette\@versim<}}
\def\ga{\mathrel{\mathpalette\@versim>}}
\def\@versim#1#2{\vcenter{\offinterlineskip
	\ialign{$\m@th#1\hfil##\hfil$\crcr#2\crcr\sim\crcr } }}

\usepackage{graphicx}
\usepackage{amssymb}
\usepackage{ociwsymp1e}
\HeadText{Juntai Shen and J. A. Sellwood}

\begin{document}

\pagenumbering{arabic}

\author[]{JUNTAI SHEN and J. A. SELLWOOD
\\Department of Physics and Astronomy, Rutgers University}

\chapter{The Survival of Bars with Central \\ Mass Concentrations}

\begin{abstract}
Many barred galaxies today harbor massive concentrations of gas in their centers, and some are known to possess supermassive black holes (SBHs) and their associated stellar cusps. Previous theoretical work has suggested that a bar in a galaxy could be dissolved by the formation of a mass concentration in the center, although the precise mass and degree of central concentration required is not well-established. Here we report an extensive study of the effects of central masses on bars in high-quality $N$-body simulations of galaxies. We study models containing both strong and weak bars, and quantify the change in bar amplitude as a mass concentration is grown in their centers. We find that bars are more robust than previously thought. We have varied the growth rate of the central mass, its final mass and the degree of concentration of the mass. Our main conclusions are: (1) the central mass has to be as large as several percent of the disk mass to completely destroy the bar; (2) for a given mass, objects whose scale-length is a few pc or less cause the greatest reduction in bar amplitude, while significantly more diffuse objects have a lesser effect; and (3) the bar amplitude always decreases as the central mass is grown, and continues to decay thereafter on a cosmological timescale. Thus the masses of SBHs are probably too small, even when dressed with a stellar cusp, to affect the bar amplitude significantly.

\end{abstract}

\section{Introduction}
\label{sec:intro}
Bars are a common component of disk galaxies; e.g.\ Eskridge \etal\ (2000) found that more than two thirds of disk galaxies in the Ohio State University Bright Spiral Galaxy Survey are either strongly or weakly barred in the near-infrared band.

Central massive concentrations (CMCs) are also frequently found in barred galaxies.  For our purposes, a CMC is any sufficiently massive object, regardless of its nature, which is likely to have a dynamical effect on the
evolution of its host galaxy.  Examples include: a molecular gas concentration with scales of $0.1\sim1$ kpc and masses of $10^7\sim 10^9M_\odot$ in the central region (Sakamoto \etal\ 1999, Regan \etal\ 2001, etc.), which is presumably a consequence of bar-driven gas inflow; a supermassive black hole
(SBH) with a mass $10^6\sim 10^9M_\odot$, or some $0.1\%M_{\rm bulge}$; a dense star cluster found near the centers of many spiral galaxies (Carollo; Walcher, this volume), which are young, extreme compact ($\leq 5$pc) and relatively massive ($10^6 \sim 10^7 M_\odot$).

Previous studies, many based on single-particle dynamics in a rotating bar potential with a CMC (Hasan, Pfenniger, \& Norman 1993 etc.), have suggested that a CMC weakens or dissolves the bar, yet CMCs appear to coexist with bars in many spiral galaxies.   In order to determine whether this presents a genuine paradox, we need to know how massive a CMC is needed to destroy the bar completely and on what timescale.

Fully self-consistent $N$-body simulations offer the best way to answer these questions.  Unfortunately, there are major discrepancies between the results reported in different publications to date.  The simulations by Norman, Sellwood \& Hasan (1996) showed that a 5\% mass can cause the bar to dissolve on a dynamical timescale.  Those by Friedli (1994), which included both stars and gas, indicated that objects with 2\% of the disk mass can dissolve the bar within about one Gyr or so.  Hozumi \& Hernquist (1998), who employed a 2-D Self Consistent Field (SCF) method, found that CMCs with 0.5 -- 1\% of the total disk mass are sufficient to weaken the bar substantially within 1 -- 4Gyr.  Also Berentzen \etal\ (1998) found, from a single experiment, that the gas inflow driven by the stellar bar leads to the formation of a massive concentration with the mass of about $1.6\%M_{\rm galaxy}$, which causes the the bar strength to decay on a timescale of 2 Gyr.   There are a few reasons to regard these results and their implications as tentative.  The massive gas content of the central regions of some barred galaxies (Sakamoto \etal\ 1999, Regan \etal\ 2001) suggests that the destruction of the bar by a CMC is less rapid than claimed.  Secondly, these large discrepancies are probably warning signs of numerical problems with the simulations or misunderstood implications.  Moreover, none of the previous studies was {\em systematic} in the sense that they did not explore the parameter space relevant to the evolution of bars in sufficient detail.

Our understanding of bar-forming mechanisms and secular evolution of barred galaxies are seriously hampered by our inadequate knowledge of the influence of CMCs on the bar.  The main motivation of this work is to confirm (or otherwise) previous work, and to give unambiguous answers to the timescale on which the bar is weakened by a CMC, the critical mass of the CMC which causes rapid bar dissolution and other parameters that affect this bar-weakening process.

\section{Model and Simulation Details}
\label{sec:model}

We construct rapidly tumbling initial bars in $N$-body experiments.  In this work we create two initial bars in different ways, which we distinguish by their relative strength.  The weak initial bar is developed by simply evolving a Kuz'min-Toomre (K-T) disk (Binney \& Tremaine 1987, \S2.2) of scale length $a$, initial thickness $0.05a$ and initial $Q=1.5$.  Some time after the bar has formed and thickened, we obtain a disk containing a long-lived bar having moderate-strength, which we define as our ``weak initial bar.''  Our ``strong initial bar'' is developed from the weak bar by adding fresh particles on circular orbits in the midplane of the outer disk in the manner described in Sellwood \& Moore (1999); the bar strength is increased for certain addition rules.

We adopt $a$ and $M$ as our units of length and mass, respectively, and our time units are therefore dynamical times $\sqrt{a^3/GM}$.  From here on all quantities are expressed in units such that $G=M=a=1$ unless otherwise noted.  These units can be scaled to physical values as desired; we adopt $M = 5\times 10^{10}M_\odot$ and $a = 3\;$kpc, which imply a unit of time of roughly $1.2\times 10^7$yr.

For simplicity, we mimic a CMC as a Plummer sphere; \ie its potential is:
\begin{equation}
\Phi_{\rm CMC}(r)=-\frac{GM(t)}{\sqrt{r^2 + \ecmc ^2}}
\end{equation}
where $\ecmc$ is the scale-length of the CMC (aka.\ the softening length or the compactness of the CMC) and $M(t)$ is the CMC mass at time $t$.  We regard $\ecmc$ as a physically interesting parameter in this study: we mimic a SBH as a ``hard'' CMC with a very small $\ecmc$ ($\sim$ a few pc, much less than the influence radius of a SBH with about one percent of galactic mass) while a ``soft'' CMC ($\ecmc\sim$ a few hundred pc) represents a massive central gas concentration. 

We grow the CMC by increasing its mass according to the relation
\begin{equation}
M(t)=\left\{
  \begin{array}{ll}
    M_{\rm CMC}\cdot \sin^2 \left( \tau \cdot \pi/2  \right) & \quad {\rm for}\quad 0 \le \tau \le 1,\\
    M_{\rm CMC} & \quad {\rm for}\quad \tau > 1
  \end{array}\right.  \label{eqn:bhmass} 
\end{equation}
with $\tau=(t-t_{\rm CMC})/t_{\rm growth}$.  This sinusoidal growth is almost exactly same as the cubic-type growth adopted in Merritt \& Quinlan (1998) and Hozumi \& Hernquist (1998) and others.

In most of our simulations we adopt a rigid dark matter halo with the potential
\begin{equation}
\label{eqn:halopotential}
\Phi_{\rm halo}= \frac{V_0^2}{2}\ln \left( 1+ \frac{r^2}{c^2} \right)
\end{equation}
which yields an asymptotically flat circular velocity of $V_0$ for $r\gg c$, with $c$ being the ``core radius''.  We set $c=30a$ and $V_0=0.7(GM/a)^{1/2}$.  We include this large-core halo as a rigid component, instead of a live one, for computational economy, and demonstrate (\S\ref{sec:tests}) that using a live halo, with very similar potential as the rigid one, gives nearly same trend
in the bar-destruction process by CMCs as for the rigid halo.

\begin{figure}[t]
\centerline{\includegraphics[angle=0, width=0.85\hsize]{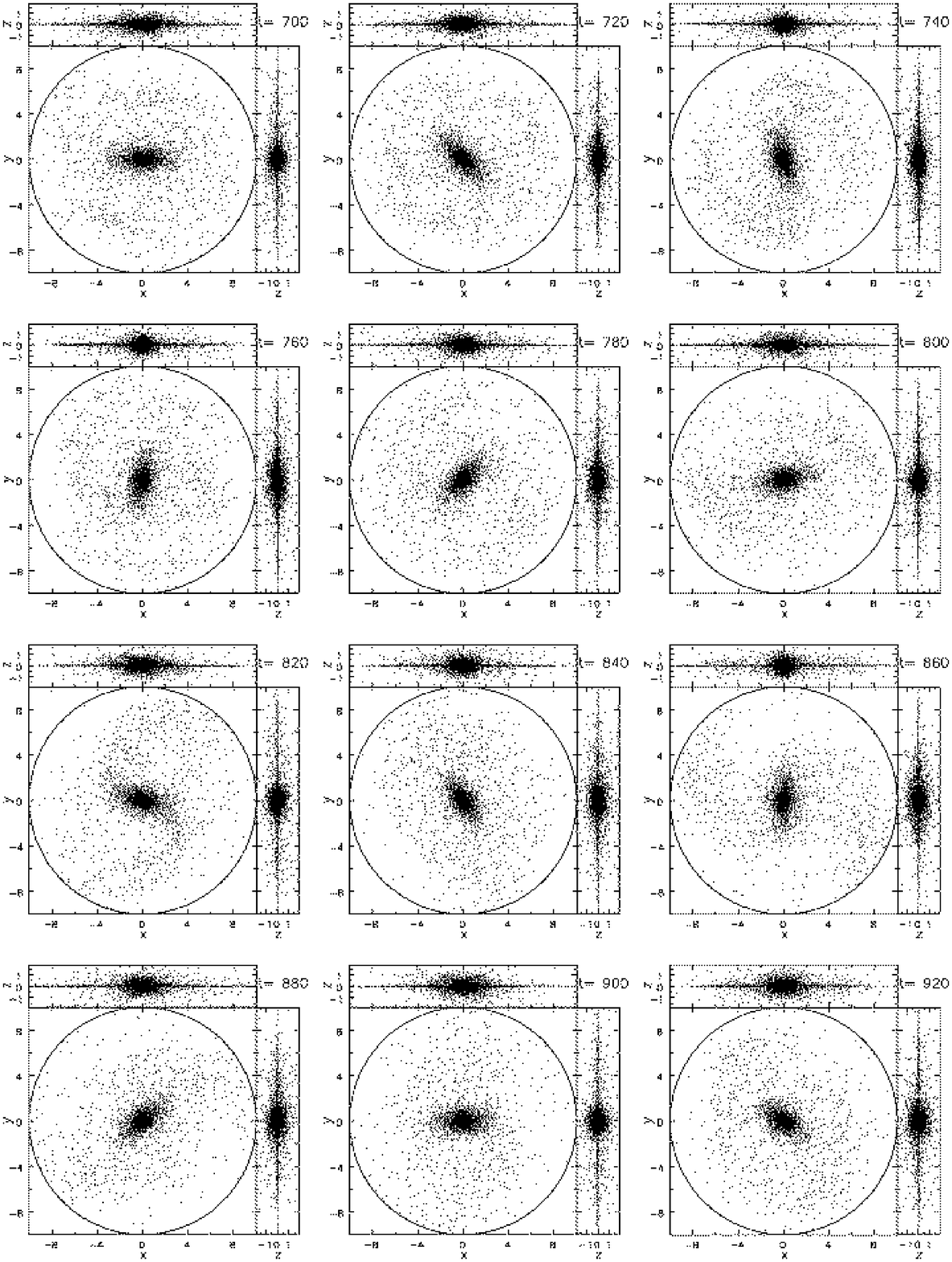}}
\caption{(a) Snapshots of particle positions showing the evolution of one simulation.  The CMC with $\mcmc=2\%$ is grown from $t=700$ to 750 according to Eq.~(\ref{eqn:bhmass}).  Only about 1 out of 500 particles is plotted and the grid extends farther vertically than shown.}
\label{fig:snapshots}
\end{figure}

\addtocounter{figure}{-1}

\begin{figure}[t]
\centerline{\includegraphics[angle=0, width=0.85\hsize]{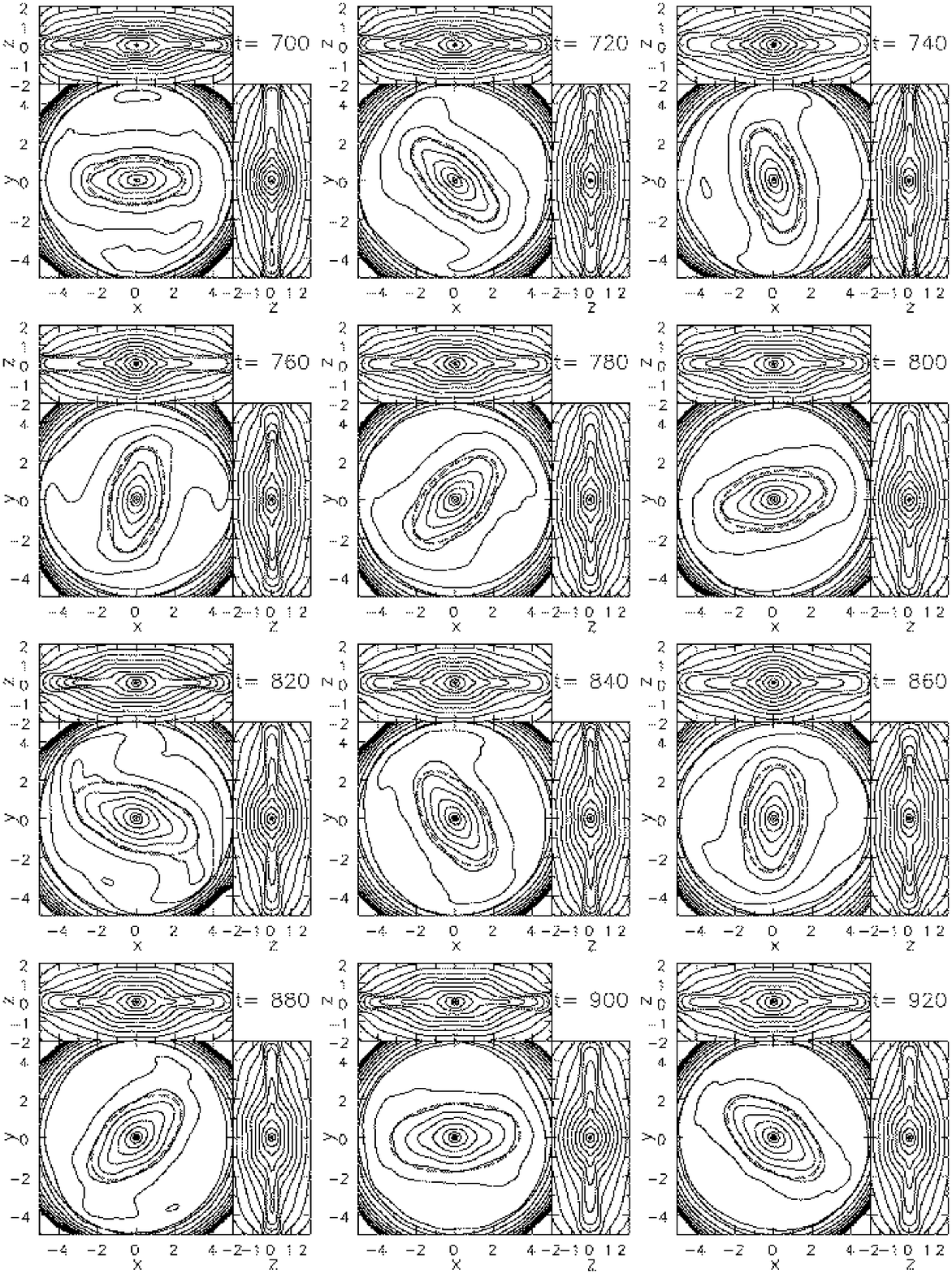}}
\caption{(b) The iso-density contours corresponding to the snapshots in (a). The contours are obtained by smoothing the discrete points with an adaptive kernel (Silverman 1986). The dashed ellipse (red) in each panel is the best fit ellipse with the largest ellipticity over SMA, analyzed with IRAF's isophote-fitting task {\tt ellipse}. These best-fit ellipses appear to match the neighboring density contours quite well. Contours are separated by a constant factor of $10^{0.4}$.}
\end{figure}

We employ the 3-D cylindrical-polar particle-mesh (PM) code described in detail by Sellwood \& Valluri (1997).  We solve separately for the three components of the gravitational force field and for the potential of the mass distribution using Fast Fourier Transforms (FFTs) in the vertical and azimuthal directions and by direct convolution in the radial direction.  The gravitational field at a distance $d$ from a unit mass particle follows the standard softened potential $\Phi(d)=-G/\sqrt{d^2+\epsilon^2}$ where the particle softening length, $\epsilon$, is a constant.  Our grid has 55 radial, 64 azimuthal and 375 vertical nodes, the particle softening length $\epsilon =0.02a$.  We generally employ more than one million particles and adopt a time step of $\dt=0.04$ for runs with no CMCs and a time step of $\dt=0.01$ when CMCs are included.  In the latter case, we implement a special ``guard annuli'' (GA) scheme (Shen \& Sellwood 2003) to ensure the motions of the most rapidly moving particles are integrated accurately.  We divide the central region around a CMC into many concentric annuli and halve the time step in each annulus as we step closer to the CMC; the shortest step can be as small as $\dt/2^9$ for hard CMCs.  Since the CMC dominates in this region, we integrate the orbit for these sub-timesteps in a fixed field, and update the self-consistent part from the bar and the disk at the basic time step interval.  For the test runs with a live halo, we use a hybrid PM scheme (Appendix B of Sellwood 2003), in which the self-gravity of the disk is computed on a high-resolution cylindrical polar grid while that of the halo is computed using a surface harmonic expansion on a spherical grid.

\section{Results}
\label{sec:results}

\subsection{A fiducial run}
\label{sec:fiducial}

\begin{figure}[t]
\centerline{\includegraphics[angle=-90, width=0.7\hsize]{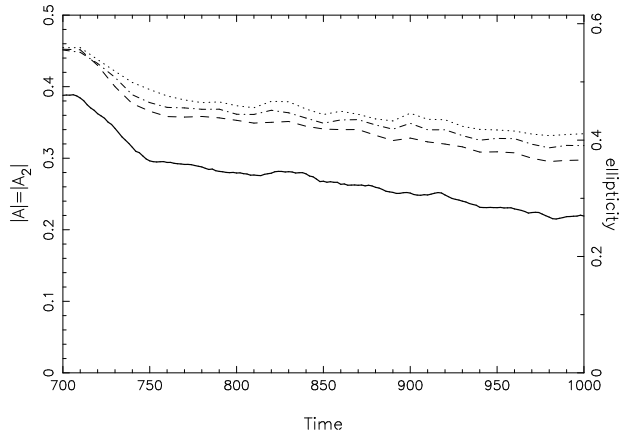}}
\caption{The time evolution of the bar amplitude $|A_2|$ (heavy solid line) and ellipticity $e$ measured at SMA=2.0, 1.75 and 1.5 (dotted, dash-dotted and dashed curves), respectively.}
\label{fig:ell_t}
\end{figure}

Figure~\ref{fig:snapshots} illustrates how a typical barred model reacts as a CMC with mass $\mcmc=2\%\mdisk$ and the scale-length $\ecmc=0.001$ is grown according to Eq.~(\ref{eqn:bhmass}) with $\tg=50$.  Figure~\ref{fig:snapshots}a shows snapshots of particle positions and Figure~\ref{fig:snapshots}b the corresponding iso-density contours.

The strong bar at time 700 rotates steadily, with period $\sim 50$, at constant amplitude in a companion run in which no CMC is grown.  In the run illustrated, the mass of the CMC rises from 0 to $2\%\mdisk$ according to Eq.~(\ref{eqn:bhmass}) between $t=700$ and 750.  The bar is weakened, but not severely damaged, by the CMC and retains a moderate strength as late as $t=920$ (Figure~\ref{fig:snapshots}).  The mass of the hard CMC used in this model is absurdly large for a central SBH, but the outcome illustrates that bars can survive even with an extraordinarily massive SBH.

Figure~\ref{fig:ell_t} shows the time evolution of two estimators of bar strength.  The heavy solid line represents $|A(t)|$, the amplitude of the $m=2$ component defined as $A(t)=A_2(t) \equiv A_{m=2}/A_{m=0}=\frac{1}{N} \sum _{j=1}^{N}\exp[im\theta_j]$, where $\theta_j$ is the coordinate of each of the $N$ particles inside the bar-spanning radial range in the simulation at time $t$.  The amplitude of the bar decreases between $t=700$ and 750, corresponding to the CMC growth phase, and much more slowly thereafter.  We also show the time evolution of ellipticity $e$, another frequently-used bar strength indicator, measured at different semi-major axes (SMAs) from IRAF isophote-fitting analysis.  The striking similarity of the overall evolutionary trends of these two estimators assures us that $|A_2|$, which is much easier to compute, is a good indicator of bar strength.

\subsection{Bar amplitude \mbox{\boldmath $|A|$} vs. \mbox{\boldmath $\tg$}}
\label{sec:A-tgrowth}
\begin{figure}[t]
\centerline{\includegraphics[angle=-90, width=0.7\hsize]{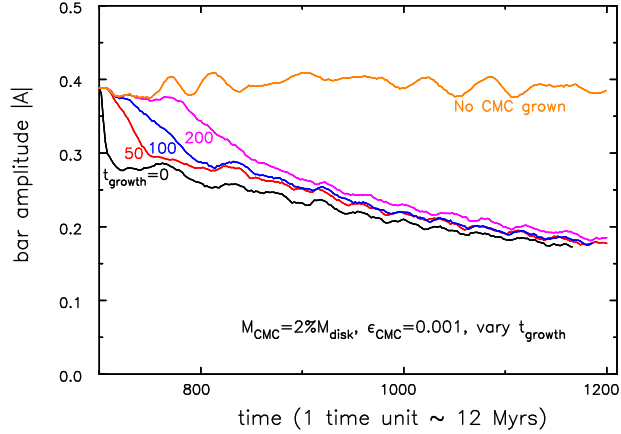}}
\caption{The bar amplitude $|A|$ evolution for runs with the same CMC, but grown with different growth time $\tg$. The uppermost curve (orange) is a comparison run with no CMC grown. Note that the pattern speed of this initial bar is about 50 time units. }
\label{fig:A-tgrowth}
\end{figure}

Figure~\ref{fig:A-tgrowth} shows how the bar amplitude evolves in runs with different $\tg$.  The dashed line shows that the bar amplitude stays roughly constant in a comparison run in which no central mass was grown.  The bar is weakened as the central mass grows, but after the CMC mass reaches its maximum value, the bar amplitude $|A|$ decays on a much longer timescale.  However, for large $\tg$ ($\ga 200$) the transition between the two trends becomes less sharp, as can be expected.  The Figure also confirms that that $\tg$ is not very important in determining the final bar amplitude long after CMC growth, as found in other studies listed in \S\ref{sec:intro}.

\begin{figure}[t]
\centerline{\includegraphics[angle=-90, width=0.7\hsize]{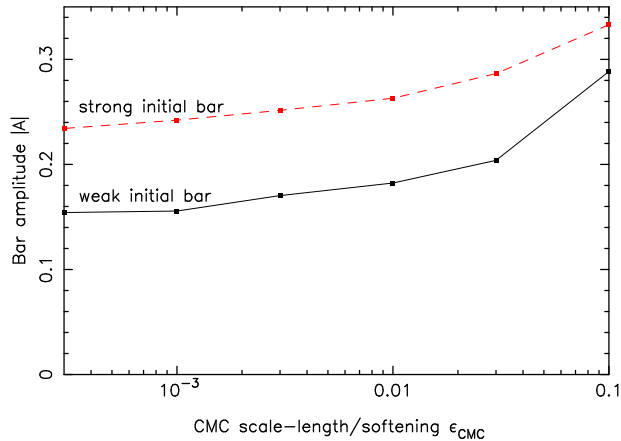}}
\caption{The bar amplitude as a function of $\ecmc$, with other CMC parameters fixed. The bar amplitude is measured 250 time units after the central mass starts to grow. The solid (dark) and dashed (red) curves represent the runs for the weak and strong initial bars, respectively. Both curves show similar trends: smaller CMC scale-length cause significantly more damage to the bar, but this effect converges for some sufficiently small $\ecmc$.}
\label{fig:A-ecmc}
\end{figure}

\begin{figure}[t]
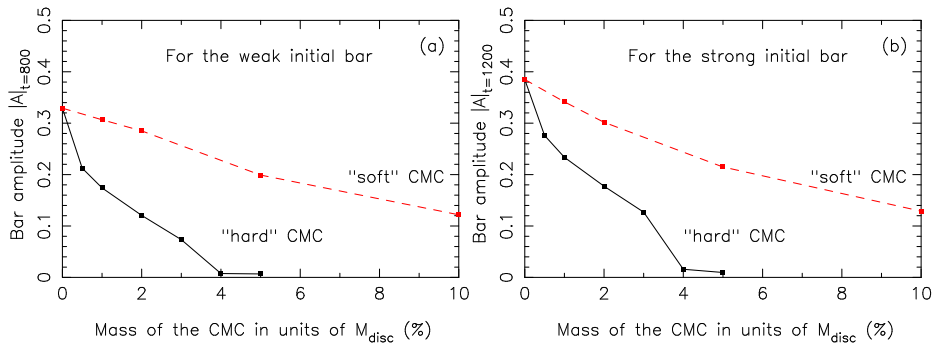

\centerline{\hspace{-0.05\hsize}\includegraphics[angle=-90, width=0.53\hsize]{w_Af_BH.ps}\includegraphics[angle=-90, width=0.53\hsize]{s_Af_BH.ps}}
\caption{(a) The final bar amplitude as a function of $\mcmc$, with other CMC parameters fixed, for the weak initial bar. The solid (dark) and dashed (red) curves represent the runs with a ``hard'' ($\ecmc=0.001$) and ``soft'' CMC ($\ecmc=0.1$), respectively. The final bar amplitude decreases continuously as $\mcmc$ is increased, and ``hard'' CMCs apparently cause significantly more damage to the bar than ``soft'' ones. A few percent $\mdisk$ for ``hard'' CMCs or more than ten percent $\mdisk$ for ``soft'' ones is needed to destroy the bar effectively on short timescales. (b) As for (a), but for the strong initial bar.}
\label{fig:A-mcmc}
\end{figure}

\subsection{Bar amplitude \mbox{\boldmath $|A|$} vs. \mbox{\boldmath $\ecmc$}}
\label{sec:A-ecmc}

The importance of the ``compactness'' of CMCs is explicitly shown for the first time in Figure~\ref{fig:A-ecmc}.  The dark and light curves show the bar amplitude, measured at a fixed time long after the CMC starts to grow, for the weak and strong initial bars, respectively.  Both types of initial bar show similar trends: {\em hard CMCs are much more destructive to bars than are those significantly more softened}.  Note that the amplitude becomes more nearly constant as $\ecmc$ decreases, which suggests convergence beyond a certain sufficiently small $\ecmc$.

\subsection{Bar amplitude \mbox{\boldmath $|A|$} vs. \mbox{\boldmath $\mcmc$}}
\label{sec:A-mcmc}
Figure~\ref{fig:A-mcmc} shows the ``final'' ($\sim 6$ Gyr after CMC growth) bar amplitude in simulations as a function of $\mcmc$, for both types of initial bars.  The final bar amplitude decreases continuously as $\mcmc$ is increased, and again we see that hard CMCs are significantly more destructive than soft ones (Figure~\ref{fig:A-ecmc}).  Independent of the initial bar strength, we find that the bar is effectively destroyed when the mass of a hard ($\ecmc=0.001$) CMC is greater than $\sim 4 $ or $5\% \mdisk$, whereas soft ($\ecmc=0.1$) CMCs, even as massive as $\sim 10 \% \mdisk$, do not totally destroy the bar within $\sim 6$ Gyr.

For the same mass, a soft CMC causes much less damage to a bar than does a hard CMC.  Since the typical sizes of molecular gas concentrations in centers of galaxies range from a few hundred pc to a few kpc (Sakamoto \etal\ 1999, Regan \etal\ 2001, etc.), corresponding to 0.1 -- 1 in our simulation units, it is clear that gaseous CMCs are less damaging than SBHs of the same mass. 

\section{Parameter Tests and Numerical Checks}
\label{sec:tests}
Simulations to study bar-destruction by CMCs are very challenging and push the limits of what is feasible with current technology and algorithms.  Here we outline a few of our many tests to check that our main results and conclusions are reliable.  Some of our findings might account for the discrepant results obtained by other groups.

We have verified that the number of particles $N$, grid size and particle softening we adopted are sufficient to reveal the correct behavior of the bar-destruction process by CMCs: variations by a factor of two or more around our adopted values make negligible difference to our main results.

\begin{figure}[t]
\centerline{\includegraphics[angle=-90, width=0.6\hsize]{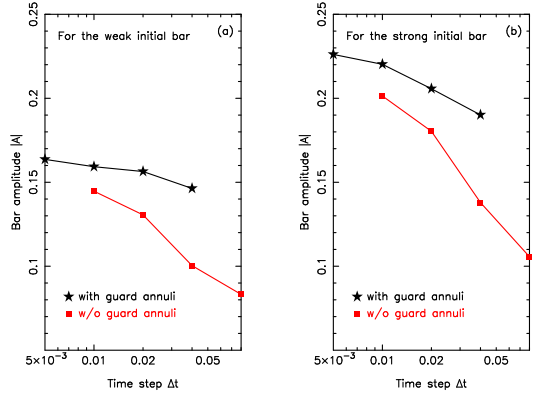}}
\caption{(a) The bar amplitude of the same test run as a function of the time step $\dt$ adopted, for the weak initial bar. The bar amplitude is measured at some fixed time after the CMC growth stopped. The filled stars (dark) and filled square (red) represent the test runs with the guard annuli scheme and with only the normal time step, respectively. The final bar amplitude might be erroneously found to be weaker than it should be as the time step gets cruder, if without special cares around a large hard CMC, like guard annuli scheme we devised. (b) As for (a), but for the strong initial bar.}
\label{fig:timesteptest}
\end{figure}

We have found the time step $\dt$ to be a very important numerical parameter; faster, but erroneous, bar destruction by CMCs occurs when orbit integration is not handled with sufficient care.  Figure~\ref{fig:timesteptest} shows that the bar amplitude at a fixed time becomes weaker as the time step is increased without employing guard annuli; there is no indication of convergence even for very small $\Delta t$.  The bar amplitude is always greater when guard annuli are active, and is less sensitive to the particular choice of the basic $\Delta t$, since we divide it even more finely as $\Delta t$ is increased.  We suspect this could be a very important factor in accounting for the faster bar decay claimed by some groups.

\begin{figure}[t]
\centerline{\includegraphics[angle=-90, width=0.7\hsize]{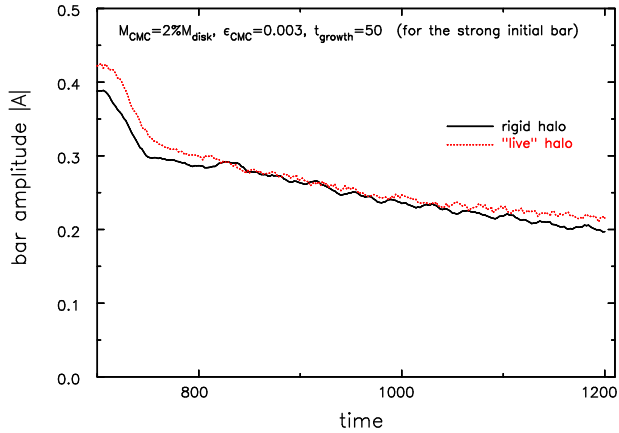}}
\caption{The comparison between test runs using a rigid and a ``live'' halo for the nearly same halo potential. The bar-weakening trends of the two runs are very similar, suggesting the use of rigid halo is not a problem for the isothermal halo potential described in Eq.~(\ref{eqn:halopotential}). And the bar destruction time scales are very similar in both rigid and live halo simulations.}
\label{fig:livehalotest}
\end{figure}

Figure~\ref{fig:livehalotest} shows that the bar amplitude evolution is hardly affected when the rigid halo is replaced by a similar live pseudo-isothermal halo with a large core resembling our rigid form (Eq.~\ref{eqn:halopotential}).  However, we also find that a {\em dense} or {\em cuspy} live halo stimulates the growth of a bar, as reported by Debattista \& Sellwood (2000) and Athanassoula (2002).  Thus, a bar in a denser or cuspy halo is {\em even harder} to destroy, which underscores one of our main conclusions that bars are extremely robust.

\section{Conclusions}
\label{sec:conclusions}

We report a systematic study of the effects of central massive concentrations (CMCs) on bars using high-quality $N$-body simulations.  We have shown that our main findings are insensitive to most numerical parameters in our simulations, but the time step requires special care.  A detailed description of our tests, results and the mechanism for bar-weakening by CMCs is presented in Shen \& Sellwood (2003).

The bar strength always decreases as the central mass is grown, and continues to decay thereafter on a very long timescale (\eg $\ga$ 6Gyr for a SBH-type CMC with 2\% $\mdisk$).  We have shown for the first time that, for a given mass, hard CMCs (SBH-type) are much more destructive to bars than soft CMCs (such as molecular gas concentrations).  We also find that bars are much more robust than previously thought; the central mass has to be as massive as a few percent of the disk mass to destroy a bar completely within a Hubble time, even for the most destructive hard CMC.

Neither typical SBHs in spirals ($\mcmc\sim 0.1\%M_{\rm Bulge}$) nor typical central molecular gas concentrations (mass $\mcmc \la 5\% \mdisk$, scale $R \sim$ a few hundred pcs) found in galactic centers should weaken the bar significantly within a Hubble time -- the former are generally are not massive enough, whereas the latter are too diffuse.  Thus, our results can naturally account for the coexistence of CMCs and bars in many spiral galaxies.  Conversely, the apparent survival of bars in galaxies with massive gaseous CMCs strongly supports our finding that gas concentrations are too diffuse to weaken bars significantly.

{\bf Acknowledgments}. This work was supported by NSF grant AST-0098282.

\end{document}